\documentclass[review,5p,twocolumn]{revtex4-1}

\usepackage{lineno,hyperref}
\usepackage{amsmath,amssymb,latexsym}
\usepackage{wrapfig}
\usepackage{subfig}
\usepackage{times}
\usepackage{epsfig}
\usepackage{graphicx}
\usepackage{color}
\usepackage{dcolumn}
\usepackage{bm}
\usepackage{setspace}
\usepackage{enumitem}
\usepackage{caption}
\usepackage{siunitx}
\usepackage[normalem]{ulem}

\usepackage{mathtools} 

\newif\ifbw
\bwfalse

\usepackage{hyperref}
\usepackage[capitalise]{cleveref}
\usepackage{amsmath}
\usepackage[toc,page]{appendix}
\usepackage{float}
\usepackage{subfig}
\usepackage{graphicx}
\usepackage{verbatim}
\usepackage{cancel}
\usepackage{color}
\usepackage[]{algorithm2e}
\usepackage{listings}

\newcommand \pxpy[2] {\frac{\partial #1}{\partial #2}}

\definecolor{dkgreen}{rgb}{0,0.6,0}
\definecolor{gray}{rgb}{0.5,0.5,0.5}
\definecolor{mauve}{rgb}{0.58,0,0.82}

\crefname{lstlisting}{algorithm}{algorithms}
\Crefname{lstlisting}{Algorithm}{Algorithms}

\begin{document}

\title{Efficient and quantitative phase field simulations of polycrystalline solidification using a vector order parameter}
%
\author{Tatu Pinomaa$^1$}
\author{Nana Ofori-Opoku$^2$}
\author{Anssi Laukkanen$^1$}
\author{Nikolas Provatas$^3$}

\affiliation{
$^1$ ICME group$,$ VTT Technical Research Centre of Finland Ltd$,$ Finland\\
$^2$ Computational Techniques Branch, Canadian Nuclear Laboratories, Chalk River, Canada\\
$^3$Department of Physics and Centre for the Physics of Materials$,$ McGill University$,$ Montreal$,$ Canada 
}

%
%
\begin{abstract}
A vector order parameter phase field model derived from a grand potential functional is presented as a new approach for modeling polycrystalline solidification of alloys. In this approach, the grand potential density is designed to contain a discrete set of finite wells, a feature that naturally allows for the growth and controlled interaction of multiple grains using a single vector field. We verify that dendritic solidification  in binary alloys follows the well-established quantitative behavior in the {\it thin interface}  limit. In addition, it is shown that grain boundary energy and solute back-diffusion are quantitatively consistent with earlier theoretical work, with grain boundary energy being controlled through a simple solid-solid interaction parameter. Moreover, when considering polycrystalline aggregates and their coarsening, we show that the kinetics follow the expected parabolic growth law. Finally, we demonstrate how this new vector order parameter model can be used to describe nucleation in polycrystalline systems via thermal fluctuations of the vector order parameter, a feature which has been thus far lacking from multi-phase or multi-order parameter based phase field models. The presented vector order parameter model serves as a practical and efficient computational tool for simulating polycrystalline materials. We also discuss the extension of the order parameter to higher dimensions as a simple method for modeling multiple solid phases.  
\end{abstract}
\date{\today}
\maketitle
\section{Introduction}

It is well established that the properties of materials and their performance are ultimately determined by their microstructure. One of the main hallmarks of ``microstructure" in materials such as metal alloys is their polycrystalline nature. One of the quintessential goals of materials processing is therefore to understand, predict, and control the polycrystalline structure during the solidification stage of their materials processing. In the last few decades, this task has been more and more undertaken using computational modeling based on so-called multi-phase field or multi-order parameter models \cite{Ste99,Bottger06,Ofori-Opoku10}. The approaches use a discrete set of fields to model different orientations of solids in the liquid, each driven by its own equation of motion that is coupled to diffusion of solutes and/or heat, and interacting with other grains  in some phenomenological manner designed into the equations. While such approaches have been very beneficial in pushing the frontiers of computational modeling, they  suffer from some limitations in theoretical consistency such as the incorporation of nucleation, thermal fluctuations, orientational invariance and requiring the use of many number of fields to simulate polycrystalline materials.


For accurate assessment and prediction of polycrystalline solidification, including nucleation, growth and grain merging, it is necessary for a simulation approach to have a self-consistent description of the magnitude of the growth kinetics and crystal anisotropy (capillary and kinetic), as well as a proper description of solute segregation and diffusion. Upon merger of grains, when a cohesive solid network has formed, coarsening ensues, which is dominated by grain boundary migration determined by grain boundary energetics and mobility. Various modeling methods have been developed to study solidification, predict polycrystalline structure formation, and the subsequent coarsening. 

Cellular automaton and kinetic Monte Carlo methods are two popular techniques that enable large-scale simulations of polycrystalline solidification. However, user intervention is required to properly define the growth kinetics, and these methods are prone to numerical issues and they have considerable limitations in addressing the simulation of alloys and diffusion controlled kinetics of solidification. 

The phase field (PF) methodology is typically the most widely used state-of-the-art numerical scheme to describe microstructural solidification. Various PF approaches exist for polycrystalline simulations. 
One family of models, pioneered by Kobayashi~\cite{kobayashi1998}, Granasy~\cite{granasy2002}, Warren \cite{Warren03}, and co-workers, is characterized by the use both orientation ($\theta$) and order parameter ($\phi$) fields (coined $\theta-\phi$ models) to  model solidification of any number of crystal orientations. This approach allows for an efficient representation of a large number of grain orientations. It successfully describes basic features of polycrystalline solidification: controllable grain boundary energy versus misorientation, polycrystalline growth, grain merger, and coarsening. However, these models are not without their challenges to implement numerically. 
Moreover, matched {\it thin interface} asymptotics has not been performed for this family of models, and it is likely to be challenging due to the inclusion of singular terms in the model description, e.g. $|\nabla \theta|$ gradient term in the free energy. Additionally, these models have been shown to exhibit unphysical topological defects, that have recently been, at least partially, addressed in 2D simulations~\cite{korbuly2017}. Finally, the extension to 3D of this class of models is not immediately straightforward, although some schemes have been explored \cite{pusztai2005}. 

Another family of PF models is called multi-phase field (MPF) models \cite{steinbach1999}. These deviate from traditional PF models in that the phase field $\phi$ represents the volume fraction of a phase, which implies that even the liquid is assigned a phase field. As with standard PF models, these models compute the energy of an interface, and anisotropy, by tuning appropriate gradient energy terms \cite{salama2020}. In addition, interaction energy between solid grains includes various polynomial interaction terms between phase fractions. However, MPF models can suffer from unphysical non-local adsorption effects due to a complicated free energy landscape wherein the sum of the phase fraction fields must be maintained by a Lagrange multiplier. These need to be eliminated in a somewhat ad-hoc manner by adding higher order polynomial penalties between phase fraction fields. Also, it is not clear how to integrate thermal fluctuations into the distinct phase fraction fields self-consistently. Another challenge with using MPF models is that, in principle, they nominally require as many fields as there are crystal grains, thus requiring significant computing resources to simulate  realistic materials domains. To circumvent this problem requires dynamic grain re-assignment algorithms to be implemented \cite{krill02}.

Yet another family of PF models is called multi-order parameter PF models. As the name suggests, in this class of models, a phase field $\phi$ maintains the traditional interpretation as a physical order parameter that distinguishes between ordered and disordered states of a crystal. As with MPF models, one order parameter is required for each crystal grain. An earlier approach developed by Ofori-Opoku et al. \cite{Ofori-Opoku10} used a quadratic repulsion energy term between order parameters, which is all that is required since this class of models does not suffer from non-local adsorption effects. This approach of Ref.~\cite{Ofori-Opoku10} allowed for a quantitative accounting of solid-liquid interface kinetics and qualitative description of solid-solid interfaces in metals.  A weakness with this implementation is that the grain boundary energy cannot be easily tuned, resulting in grain boundary energies corresponding to high angle/energy grain boundaries. As with MPF, for multi-order parameter models the Langevin noise based nucleation is problematic to implement when multiple order parameters are fluctuating in the same material point due to the use of polynomial (here quadratic) repulsive interaction terms.

In this work, we present an alternative phase field modeling scheme to describe solidification and coarsening of multiple crystallographic orientations in the context of a physical order parameter description. We accomplish this by introducing a two dimensional vector order parameter as in Ref.~\cite{morin1995}, which is then integrated with a previous quantitative modeling scheme for solidification based on a grand potential energy functional \cite{Plapp11}.  This main innovations of this approach are the trivial generation of a Landau free energy landscape corresponding to any number of discrete orientations in the context of a {\it single} physical order parameter field, simple  control of grain boundary energy, and self-consistent incorporation of thermal fluctuations into dynamical simulations.  

The paper is organized as follows. In \Cref{sec:methods}, we introduce the vector order parameter model and review its parameterization.  \Cref{sec:results} examines the model's capacity to reproduce some important benchmarks quantitatively. These include, free solidification of a single dendrite, grain boundary energy, grain merger and back-diffusion, grain growth and coarsening, grain boundary segregation in polycrystalline growth, and noise-induced nucleation. \Cref{sec:discussions} further discusses some of the qualities of our model and potential further extensions.  We conclude in \Cref{sec:conc}.

\section{Methodology}
\label{sec:methods}

This section develops the formalism of the vector phase field model. In this work, the focus is to demonstrate the properties of this phase field methodology, and therefore we limit ourselves to binary alloys. We will formulate the model formally in the grand potential ensemble, meaning that the dependent fields are the order parameter and the chemical potential. We begin by describing the grand potential energy functional, highlighting its vectorized Landau landscape. Following this, we discuss the dynamics of the vector order parameter and the chemical potential. We then highlight the parameter relations necessary to achieve quantitative correspondence with the sharp interface model of solidification.

\subsection{Grand potential functional}
\begin{figure}
\includegraphics[width=0.47\textwidth]{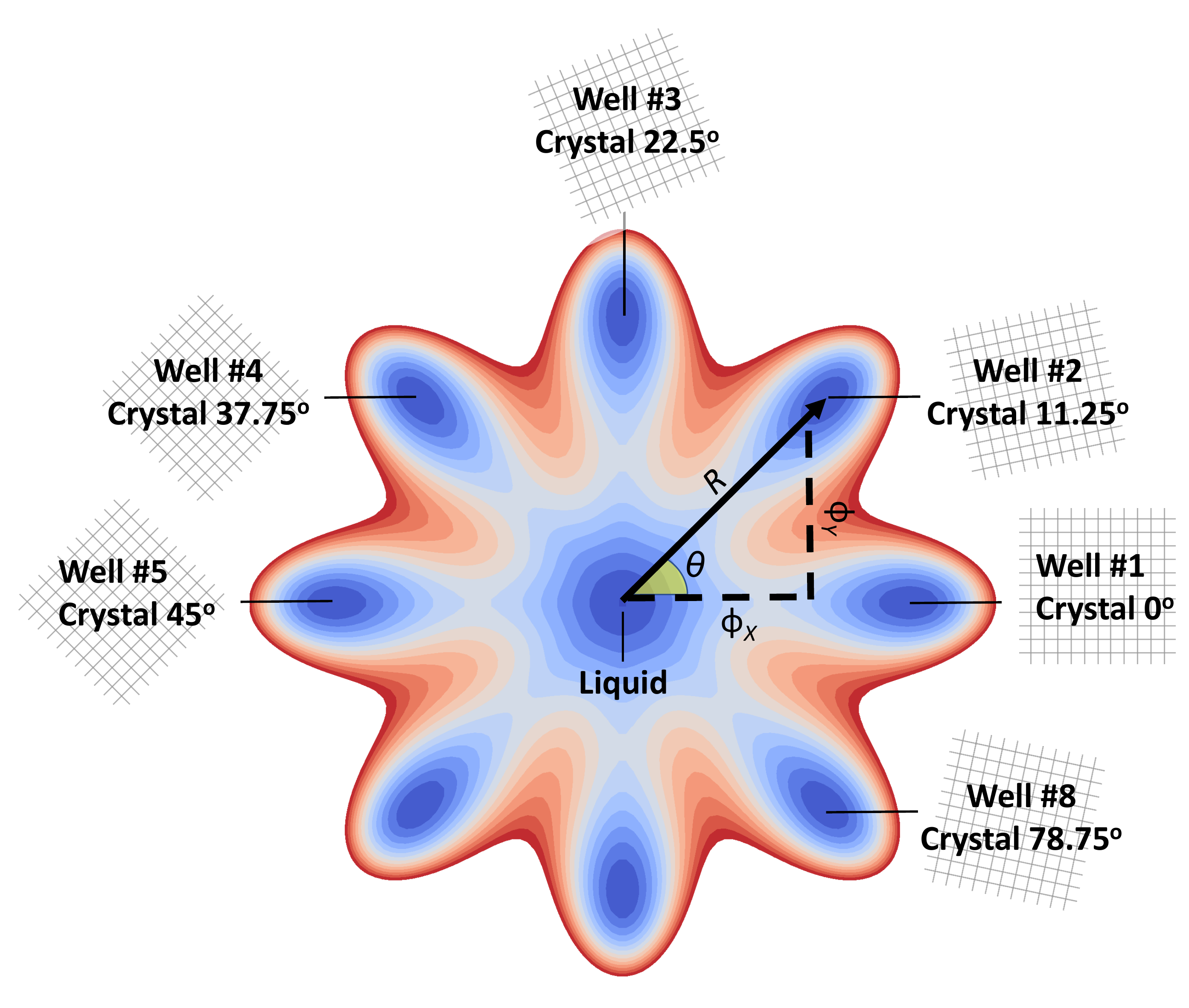}
\caption{Landau free energy landscape in the space of a vector order parameter with $N_{wells} = 8$ and a solid-solid interaction strength parameter $b = 0.05$. The energy minimum in the center corresponds to the liquid well, while each of the $N_{wells}$ correspond to a distinct solid orientation. In this work, the dynamics of the order parameter are written in term of its Cartesian projections $\phi_X = R \cos(\theta)$ and $\phi_Y = R \sin(\theta)$. 
}
\label{fig:landau_landscape}
\end{figure}

We start by defining a vector order parameter defined as  $\vec{\phi}(\vec{x},t)=(\phi_X(\vec{x},t),\phi_Y(\vec{x},t ))$. We also project $\vec{\phi}$ into polar coordinates $R$ and $\theta$ according to
\begin{align}
    \phi_X &= R \cos(\theta)
    \\
    \phi_Y &= R \sin(\theta),
    \label{phi_to_R_theta_projection}
\end{align}
where $R$ can be interpreted as the ``traditional'' solid-liquid order parameter, and $\theta$ is an additional degree of freedom that is used to design a Landau functional whose bulk part contains a discrete number of solid wells. This is given by 
\begin{align}
     f_{\rm{Landau}}(R,\theta) = R^2 - 2\frac{1+b \, \cos\left( N_{wells} \theta \right)}{1+b}R^3 + R^4,
     \label{landau_bulk_part}
\end{align}
where $N_{wells}$ sets the number of discrete solid wells corresponding to unique crystalline orientations, and $b$ controls the free energy barrier between neighboring solid wells. \cref{landau_bulk_part} is usually referred to as the ``double well function" in conventional PF models, where there is one solid well and one liquid well.  An example of the Landau landscape is shown in \cref{fig:landau_landscape} with $N_{wells} = 8$ and $b=0.05$. Our approach is analogous to Morin et al. \cite{morin1995}, who used a different order polynomial in $R$ to describe the Landau free energy function (polynomial orders $R^n$, $n=2, 4, 6$). Here, we modify their approach to a polynomial which is a natural extension of the single order parameter PF models typically used for solidification, thus adopting polynomial orders of $R^n$, $n=2, 3, 4$.
It is noted that for 2D cubic lattices, the physical crystalline orientation is one fourth of the Landau angle $\theta$. This is illustrated in \cref{fig:landau_landscape}. Thus, for example, $\theta = \pi$ (180$^o$) corresponds to the maximum crystal misorientation, which for a cubic 2D lattice is $\pi/4$ (45$^o$). 

To the multi-well function in Eq.~(\Ref{landau_bulk_part}) we add a chemical energy contribution of the form 
\begin{align}
     \omega_{\rm{ch}}(\mu,R) = P(R) \,
    \omega^{s}(\mu,T) + \Big(1-P(R) \Big)\,\omega^\ell(\mu,T),
     \label{gp_bulk_part}
\end{align}
where $\mu=\mu(\vec{x},t)$ is the chemical potential field, $\omega^{s}$ and $\omega^\ell$ are the grand potential densities of the solid and liquid phases, respectively, and $P(R)$ interpolates the thermodynamic potentials across bulk states. It satisfies $P(0)=0$ and $P(1)=1$.  \cref{gp_bulk_part} provides the chemical driving force for solidification written in terms of the chemical potential, which is the natural thermodynamic variable of the grand potential  functional.

The last contribution to the grand potential energy requires gradient energy terms to account for non-local interactions. In terms of the Cartesian representation of the order parameter, we also define gradient interaction term as according to 
\begin{align}
f_{\rm{int}} = \frac{1}{2}W_{X}^2|\nabla \phi_X|^2 + \frac{1}{2}W_{Y}^2|\nabla \phi_Y|^2,
\label{eq:fint}
\end{align}
where $\lbrace W_X, W_Y \rbrace$ can be shown to be proportional to gradient energy and inversely proportional to the square root of the nucleation barier \cite{ProvatasElder10}. The coefficients must be made anisotropic in order to self-consistently reproduce the symmetries of crystal growth in solidification. The anisotropy of these coefficients will be discussed further below. 

Combining Eqs.~(\ref{landau_bulk_part}), (\ref{gp_bulk_part}) and (\ref{eq:fint}) yields a grand potential functional of the form
\begin{align}
    \frac{1}{H}\Omega[\vec{\phi},\mu,T]  =
    \int_V & \Bigg( 
    \frac{1}{2} W_X^2 |\nabla \phi_X|^2 +
    \frac{1}{2} W_Y^2 |\nabla \phi_Y|^2
    \Bigg. \nonumber \\ \Bigg.
    +& R^2 - 2\frac{1+b \, \cos\left( N_{wells} \theta\right)}{1+b}R^3 + R^4
    \Bigg.  \nonumber \\ \Bigg. 
    +& \frac{1}{H} P(R) 
    \omega^{s}(\mu,T) 
    \nonumber \\
    +& \frac{1}{H} \Big(1-P(R) \Big) \omega^\ell(\mu,T)
    \Bigg) dV, 
    \label{eq:totalFreeEnergyFunctional}
\end{align}
where $H$ is the nucleation barrier, $T$ is the temperature and $P(R)$ is a function that interpolates the grand potential between its bulk solid to liquid values. An explicit form of $P(R)$ introduced by Wheeler-Boettinger-McFadden \cite{WBM1992} is $P(R) = 10R^3 - 15R^4 +6R^5$, although several other forms are possible.

\subsection{Phase field dynamics}

The dynamical evolution  for the vector order parameter  is governed by the usual dissipative relaxation dynamics in each component field, while solute evolution follows the usual mass conservation dynamics. The explicit dynamical equations for these fields are given by 
\begin{align}
    \pxpy{\phi_K}{t}&=-\pxpy{\Omega}{\phi_K}+\eta_K, \,\, K=\{X,Y\} \label{general_dynamics_phiK}\\
     \pxpy{c}{t}&=\nabla \cdot \left( M(\vec{\phi},c) \nabla \mu\right),
    \label{general_dynamics_c}
\end{align}
where $K=\{X,Y\}$ denote the two components of $\vec{\phi}$, $\eta_K$ are noise sources in each component (discussed further below), while $c(\vec{x},t)$ is the solute concentration field of a binary alloy, which in the grand potential ensemble is determined via the chemical potential field $\mu(\vec{x},t)$ according to the relation   
\begin{align}
    c=-\frac{\delta \Omega}{\delta \mu}=P(R)c^s(\mu)+\big(1-P(R)\big)
    c^{\ell}(\mu),
    \label{GP_C_form}
\end{align}
where $c^s\!=\!\partial \omega^s / \partial \mu$ and $c^{\ell}\!=\!\partial \omega^{\ell} / \partial \mu$. \cref{GP_C_form} can be seen as a constraint between $c$ and $\mu$.  
Combining \cref{GP_C_form} with \cref{general_dynamics_c}  yields the following evolution equation for $\mu(\vec{x},t)$ directly,   
\begin{align}
\chi(\mu) \pxpy{\mu}{t} \!=\!  \nabla \!\cdot\! \left( M(\vec{\phi},c) \nabla \mu\right) 
\!-\! P^{\prime}(R)\!\left[ c^{s}(\mu) \!\!-\!\! c^{\ell}(\mu) \right] \! \pxpy{R}{t},  
\label{general_dynamics_mu}
\end{align}
where the  susceptibility field $\chi(\mu(\vec{x},t))$ is given by 
\begin{align}
\chi = P(R)\,\pxpy{c^{s}(\mu)}{\mu} + \left[1 - P(R) \right] \pxpy{c^{\ell}(\mu)}{\mu}
\label{suscep_func}
\end{align}
Either \cref{general_dynamics_mu} or \cref{general_dynamics_c} can be coupled to the order parameter component dynamcis of \cref{general_dynamics_phiK}. 

In what follows, we will be focusing on examining the efficiencies and self-consistencies afforded when using a vector order parameter in favor of multiple scalar order parameters or phase fractions in phase field modeling. As a result, we will limit the system studied to a dilute binary alloy. We will simulate phase field dynamics using Eqs.~(\ref{general_dynamics_c}) and (\ref{general_dynamics_phiK}). Furthermore, in order to achieve quantitative solidification results in the limit of a diffuse interface, we will incorporate an anti-trapping flux in \cref{general_dynamics_c}, whose role is to suppress spurious kinetic behavior cause by the use of a diffuse interface \cite{Karma01,Echebarria04}. It is noted that the dynamical evolution equation for mass transport no longer follows a variational formulation when anti-trapping is used.

Taking into account the above considerations, we proceed as follows to arrive at an evolution equation for the vector order parameter components of our model: first we substitute the grand potential densities for the solid and liquid phases of an ideal binary alloy into \cref{eq:totalFreeEnergyFunctional}; secondly, we carry out the variational derivatives as specified in the first of Eqs.~(\ref{general_dynamics_phiK}); thirdly, we simplify the chemical driving force analogously to Ref.~\cite{Plapp11}. This yields,
\begin{align}
 \tau_K  \pxpy{ \phi_K }{t}  
 =&
\nabla \cdot \left( \pxpy{f_{\rm int}}{\left(\nabla \phi_K\right)} \right) -\pxpy{f_{\rm int}}{\phi_K}
 +\left( 2 + 3R - 4R^2\right) \phi_K
 \nonumber \\&
 + 6\frac{b\left(\, \cos\left( N_{wells} \theta\right)-1\right)}{1+b}R \phi_K
 \nonumber \\&
 + 2 b N_{wells} \frac{ \sin\left( N_{wells} \theta\right)}{1+b} R \pxpy{\theta}{\phi_K}
\nonumber \\&
 - \frac{\lambda}{1-k_e} \left( e^{u} - 1 - \frac{ T_\ell - T}{|m_\ell^e| c_0} \right) (1-R)^2 \phi_K 
 \nonumber \\& + \eta_K \text{ for } K \in \lbrace X,Y \rbrace,
\label{eq:phi_K_evolution}
\end{align}
where  $u$ is a dimensionless chemical potential difference relative to the equilibrium chemical potential for a binary alloy, expressed in terms of concentration as $u = \ln{[c/(c_0\cdot(1-(1-k_e)R))]}$ \cite{ProvatasElder10}. The first two terms on the right-hand side of \cref{eq:phi_K_evolution} represent the interface energy penalty, whose expressions are rather lengthy, and are thus shown in the Appendix. $\tau_{K}$ is the anisotropic time constant, $\lambda$ is the coupling constant, which is used to control the model's convergence onto the key results of the sharp interface model and $k_e$ is the equilibrium partition coefficient. We have also tacitly appended into the chemical driving force of \cref{eq:phi_K_evolution} a temperature component, where $T_\ell$ is the liquidus temperature, $T$ is the local temperature, $m_\ell^e$ is the liquidus slope, and $c_0$ is a reference alloy concentration. The last term in \cref{eq:phi_K_evolution}, $\eta_K$, is thermal noise,  discussed further below.

The concentration evolution follows from  \cref{general_dynamics_c} and maps identically onto the  dilute binary alloy model of Refs.~\cite{Karma01,Echebarria04}, with $R$ replacing the solid-liquid order parameter, but where $R$ is scaled here to the limits $[0,1]$. Explicitly, the concentration equation becomes
\begin{align}
\pxpy{c}{t}
= 
\nabla \cdot & \Bigg[
\Bigg( h(R)\, D_s k_e
    +
    D_\ell  \Big(1-h(R) \Big) \Bigg) \nabla e^u
\nonumber \\
&  
+ W_0 \, a_t c_0(1-k_e)e^u
\pxpy{R}{t} \frac{\nabla R}{|\nabla R|} \Bigg)
\Bigg],
\label{eq:c_evolution}
\end{align}
where $h(R)=R$ is an interpolation function, $D_S$ ($D_L$) is the solute diffusion coefficient in solid (liquid), and $a_t$ is the anti-trapping current prefactor. Its value is listed in  \Cref{table:phaseField_parameters} and is determined by conducting a matched asymptotic analysis using the methods in Refs.~\cite{Echebarria04,ProvatasElder10}.

A key feature of quantitative phase field modeling is the use of matched interface asymptotic analysis to determine the parameter relations that allow the model to emulate the interfacial kinetic of the classical sharp interface model of solidification. There have been several important works on this topic for models using a scalar order parameter \cite{Karma98,Almgren99,Echebarria04,ProvatasElder10}. The outcomes are similar; they lead to relationships between microscopic phase field parameters ($\tau_0$, $W_0$ and $\lambda$) and the macroscopic material parameters of the sharp interface model, namely, the capillary length ($d_0$) and kinetic coefficient ($\beta$).

It turns out that we can directly apply the aforementioned matched asymptotic analysis methods, developed for scalar order parameter PF models, in order to inter-relate the parameters $\tau_K$, $W_K$ and $\lambda$ in our model. The rationale in doing so is as follows: if we ignore anisotropy (as is done in the the above-referenced approaches) and only consider single individual order parameter profiles going from the liquid well into any of the solid wells of our model (see \cref{fig:landau_landscape}), we can map $\tau_K \rightarrow  \tau_0$ and $W_K \rightarrow  W_0$. Further, in this situation, the Landau angle variable becomes a constant everywhere, equal to $\theta = m\, 2 \pi /N_{wells}$ where $m$ is an integer. This guarantees the elimination of the cubic terms in the Landau contribution to the energy. This reduces \cref{eq:phi_K_evolution}) identically to the traditional form of the scalar order parameter evolution equation in both order parameter components, $\phi_K, K=X,Y$. As a result, the matched interface asymptotics in Refs. \cite{Almgren99,Echebarria04,ProvatasElder10} is exactly applicable. It should be noted that this assumes that the order parameter profile can ``ride the rails'' along a line between the liquid well to any of the solid wells. This is easily satisfied, and any error is negligible in our single crystal benchmark tests presented in the \Cref{section:results_singleCrystalSolidification}. 
These considerations lead to the following relations,
\begin{align}
d_0 &= a_1 \frac{ W_0}{\lambda}
\label{eq:d_o_thinInterfaceRelation}
\\
\beta &= a_1 \frac{ \tau_0}{\lambda \, W_0 }\left[1 - a_2 \frac{ \lambda W^2_0}{ \tau_0 D_\ell}\right],
\label{eq:beta_thinInterfaceRelation}
\end{align}
where $a_1$ and $a_2$ are asymptotic analysis constants listed in \Cref{table:phaseField_parameters}, where it is notable that solid-liquid ordering is in the interval $[0,1]$, while perhaps a more common practice is to scale the order parameter for the interval $[-1,1]$. 

To make the model anisotropic, the phase field parameters $W_K$ and $\tau_K$ are appended with an anisotropy correction according to $W_K = W_0 a_K(\phi_X,\phi_Y)$, and assuming that $\beta = 0$, $\tau_K = \tau_0 a_K(\phi_X,\phi_Y)^2$ (assuming $\beta = 0$), where $a_k$ is an anisotopy function. For 2D cubic crystals, $a_k$ is defined by 
 \begin{align}
    a_K(\phi_X, \phi_Y)  =  1 + \epsilon_c \cos \left(4 \theta_{\nabla \phi_K} - \theta \right) , K \in \lbrace X,Y \rbrace,
    \label{eq:anisotropy_function}
\end{align}
where $\epsilon_c$ is the anisotropy strength. It should be noted that in \cref{eq:anisotropy_function}, the Landau angle $\theta$ is not multiplied by four to achieve the correct rotation for a cubic 2D lattice, as visualized in \cref{fig:landau_landscape} . Here, $\theta_{\nabla\phi_K}$ is the local solid-liquid interface normal angle for vector order parameter component $\phi_K$ ($K \in \lbrace X,Y \rbrace$, given by
\begin{align}
    \theta_{\nabla \phi_K} = \arctan \left( \frac{ \partial_y \phi_K}{\partial_x \phi_K} \right),\,\, K \in \lbrace X,Y \rbrace 
\end{align}
where $\theta$ sets the reference Landau angle, i.e.,
\begin{align}
    \theta = \arctan \left( \frac{  \phi_Y}{\phi_X} \right).
\end{align}

The last term, $\eta_K$ for $K \in \lbrace X,Y \rbrace$, represents  thermal fluctuations, which must satisfy the fluctuation-dissipation theorem in order to assure convergence of the phase field equations to thermodynamics equilibrium. We follow the  approach given in Ref. \cite{karma1999} to scale these noise sources quantitatively, which yields
\begin{align}
    \langle \eta_K, \eta_K \rangle
    &=  2 k_B T \frac{1}{H \tau_0} \delta(\mathbf{x}-\mathbf{x}') \delta(t-t')
    \nonumber \\
    &=  2 k_B T \frac{1}{\tau_0} \frac{a_1 W_0}{ J d_0}\, \frac{1}{ c_0^2 \,g_{c,c}^\ell}\,
    \frac{1}{\Delta x^3} \frac{1}{\Delta t}
    ,
\label{eq:eta_K_noise}
\end{align}
where $k_B$ is Boltzmann's constant. On the second line we approximate the delta functions by $\Delta x$ and $\Delta t$ following Ref. \cite{karma1999}. Furthermore, in the second line of \cref{eq:eta_K_noise}, we substituted $H$ with an expression from the matched interface asymptotics for dilute binary alloys \cite{Echebarria04,ProvatasElder10}, where $J=1/30$ is a scaling factor (specific to the interpolation function we use for the driving force), and $g_{c,c}^\ell$ is the liquid free energy curvature. In addition, the noise term in \cref{eq:eta_K_noise} is filtered in space and time so that the fluctuations occur approximately in the length scale of $W_0$ and time scale $\tau_0$, as these can be considered as the smallest physical scales in solidification that the current phase field model can resolve. More details about this filtering scheme can be found in the Supplemental material \cite{supplementalMaterial}.

\subsection{Alternate gradient formulation}
It is noted that as in Morin et al. \cite{morin1995}, we chose a gradient energy penalty in \cref{eq:fint} that is a sum of the contributing components, and our coefficients are made anisotropic to satisfy a crucial criterion required for solidification of metals.  Another perhaps more consistent choice would be to represent the energy penalty in terms of polar coordinates $R$ and $\theta$, in the form 
\begin{align}
    \frac{1}{2}W^2|\nabla R|^2 + \frac{1}{2}W^2 R^2 |\nabla \theta|^2.
\end{align} 
While this form reproduces some features of solidification and coarsening qualitatively, this $R-\theta$ representation suffers from lattice pinning due to topological defects, analogous to those seen in the orientation PF models \cite{korbuly2017}. Furthermore, it is unclear how to implement thermodynamically consistent noise for the $\theta$ field in our vector order parameter model. As such this alternate formulation is not considered further in this work and will be pursued in future work.

\section{Model Benchmarks}
\label{sec:results}
In what follows, we demonstrate that our vector order parameter model reproduces certain established benchmark results of solidification and solid-state coarsening phenomena. In particular, we first analyze single crystal dendrite growth kinetics. We then conduct an analysis of grain boundary profiles, their energies, and their coalescence. Following this, we verify that the polycrystalline coarsening for Model A dynamics (i.e. no  concentration diffusion) is consistent with theoretical expectations. Finally, we demonstrate  nucleation and growth of a polycrystalline alloy system using thermal noise fluctuations of the vector order parameter. 

\subsection{Numerical implementation}

All phase field simulations we present are performed using the computational platform presented in Ref. \cite{Greenwood18}, which contains adaptive mesh refinement (AMR) and distributed memory parallelization based on MPI.

Concentration evolution in \cref{eq:c_evolution} was implemented using a finite volume method, as well as the divergence terms in the order parameter equations of \cref{eq:phi_K_evolution}. Other gradient terms were expressed via simple five stencil finite differencing. 
Neumann, i.e. zero flux, boundary conditions were applied, unless otherwise stated. We use forward Euler time integration, where it noted that the forward Euler time step of the stochastic noise term $\eta_K$ in \cref{eq:phi_K_evolution,eq:eta_K_noise} is proportional to $\sqrt{\Delta t}$. 
\begin{table}
\caption{Phase field model parameters for Al-4.5at\%Cu.}
\begin{tabular}{lcc}
\hline
Partition coefficient $k_e$  & 0.15  
\\
Liquidus slope $m_\ell^e$   & -5.3 K/wt\%
\\
Alloy concentration $c_0$    & 4.5 at\%
\\
Gibbs-Thomson coefficient $\Gamma$   & 2.41$\times$ 10$^{-7}$ K m 
\\
Solutal capillary length $d_0$   & 12.17 nm 
\\
Liquid free energy curvature $g_{c,c}^\ell$ & 9.05$\times$10$^9$ J/m$^3$ $^a$
\\
Liquid diffusion coefficient $D_\ell$ & 4.4 $\times $10$^{-9}$ m$^2$/s 
\\
Solid diffusion coefficient $D_s$  & 0 or 4.4$\times$10$^{-11}$ m$^2$/s $^{b}$
\\
Kinetic coefficient $\beta$           & 0.0 s/m
\\
Capillary anisotropy strength $\epsilon_c$       & 0.02 
\\
Interface width $W_0$ & $d_0/0.277$
\\
Asymptotics constant $a_1$& 7.07107
\\
Asymptotics constant $a_2$& 0.078337
\\
Asymptotics constant $a_t$& 1/$\sqrt{2}$
\\
Mesh spacing $\Delta x$ & 0.1$W_0$ / 0.2$W_0$ / 0.6 $W_0$ $^{c}$
\\
Time step size $\Delta t$ & 0.8 $\Delta x^2/(6 D_\ell)$
\\
\hline
\end{tabular}
\\
$^a$: Value from Thermo-Calc TCAL database.
\\ 
$^{b}$: $D_s$ is non-zero only for back-diffusion analysis in \cref{fig:coalescence}. 
\\
$^{c}$: We use different $\Delta x$ depending on the sections: $\Delta x = 0.6 W_0$ for the single crystal solidification benchmark  ( \Cref{section:results_singleCrystalSolidification}), $\Delta x = 0.1 W_0$ for 1D grain boundary analyses (Section \ref{section:results_1Danalysis}), and $\Delta x = 0.2 W_0$ for polycrystalline simulations (Sections \ref{section:results_modelA} and \ref{section:results_modelC}).
\label{table:phaseField_parameters}
\end{table}

\subsection{Convergence analysis for dendritic solidification}\label{section:results_singleCrystalSolidification}

Since our vector order parameter model relies on the same diffuse interface analysis as Refs.~\cite{Karma01,Echebarria04}, we verified that its dynamics reproduces the dendritic solidification results predicted by the classical sharp interface model of solidification. To do so, we initialized a seed with radius $22 d_0$ in a uniform supersaturation $\Omega = c/(c_0 (1-(1-k_e)R) = 0.55$. We then performed a reference simulation using the scalar order parameter model presented in Ref.~\cite{Karma01} and compared it to a simulation conducted with our model with $N_{wells} = 8$ and a solid-solid interaction barrier parameter $b=4.0$. We set $W_0 = d_0/0.277$ and $\Delta x = 0.6 W_0$ in our simulations. 

\begin{figure}
\includegraphics[width=0.49\textwidth]{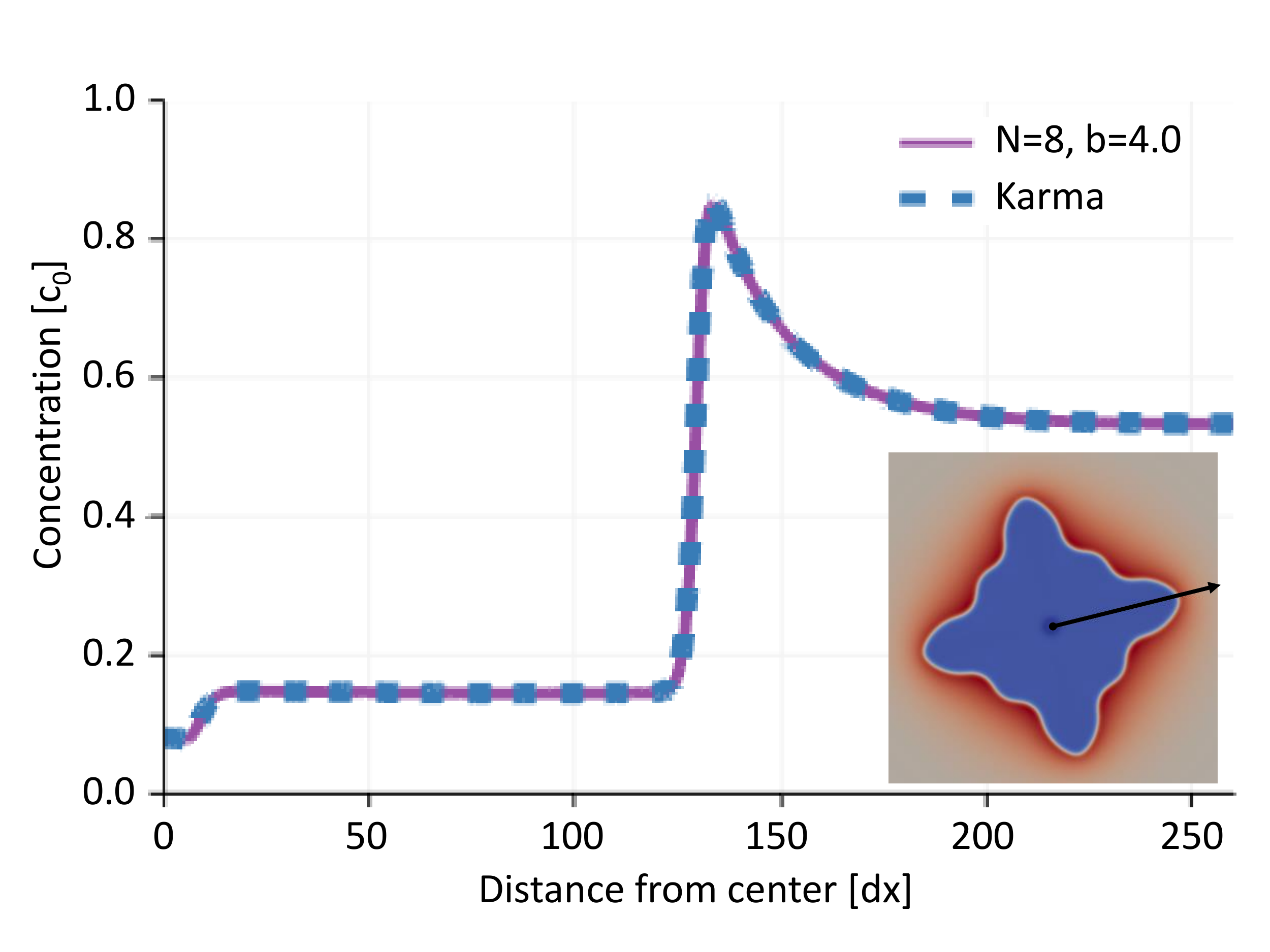}
\caption{Binary alloy dendrite growth benchmark.  Centre-line concentration computed with the scalar PF model of Ref.~\cite{Karma01} (dashed line) compared to corresponding result from the new vector order parameter model with $N_{wells} = 8$ and $b=4.0$ (solid line). In both cases, the grain is rotated 11.25$^o$, $W_0 =d_0/0.277$, the initial seed radius is $22 d_0$ and the initial supersaturation $\Omega = 0.55$. (Bottom right inset) dendrite solute contours. The profiles are recorded at time $t=480 \tau_0$. 
}
\label{fig:karma_benchmark}
\end{figure}

The solid seed is initialized into the second well of our Landau energy landscape (Well \#2 in \cref{fig:landau_landscape}), which corresponds to a crystalline rotation of 11.25$^o$ relative to x-axis. We applied the same lattice rotation to the standard phase field model of Karma \cite{Karma01}. The resulting dendrite morphology and center-line concentration through to the dendrite tip are shown in \cref{fig:karma_benchmark} at a time $t=480 \, \tau_0$. The results show excellent agreement between the standard scalar order parameter alloy model and our vector order parameter model. It is noted that these results do not change with a decrease of the solid-solid barrier parameter to $b=0.1$. We also verified that the tip speeds are indistinguishable (not shown here).
It is noted that an excessively large interaction parameter $b$ leads to unphysical oscillations inside the solid. These can be suppressed by decreasing $\Delta t$. Thus, for $N_{wells}=8$, simulations with $b=0.1$ are stable for $\Delta t=0.8  \Delta x^2/(6 D_\ell)$, while for $b=4.0$ $\Delta t=0.4 \Delta x^2/(6 D_\ell)$ is required to maintain stability.

\subsection{1D grain boundary analysis}
\label{section:results_1Danalysis}

This section analyses grain boundaries and their energy in the context of the vector order parameter model. We perform simulations in one spatial dimensional to model a 1D grain boundary analysis using so-called Model A dynamics, i.e. only the vector order parameter evolution of \cref{eq:phi_K_evolution} is considered, while the concentration dynamics in \cref{eq:c_evolution} is switched off. 

We applied a fixed value boundary condition so  that the far ends of the 1D grain boundary geometry satisfy $\phi_X^2 + \phi_Y^2 = 1$, and such that $\theta=\arctan(\phi_Y/\phi_X)$ corresponds to the appropriate crystal orientations. The grid spacing was set to $\Delta x = 0.1 W_0$, even though $\Delta x = 0.2 W_0$ also yields convergent results. It noted that the use of larger grid spacing, such as $\Delta x=0.5 W_0$, may lead to lattice pinning in the multidimensional polycrystal coarsening simulations, which are discussed in  \cref{section:results_modelA}. Grain boundary equilibration was conducted by initializing two 1D ``solid blocks" corresponding to two different solid wells, with hyperbolic tangent tails overlapping at $R=0.5$. Approximately $20 000$ time steps were required to equilibrate the profiles into a static grain boundary. The final equilibrium profiles were found to be independent of the details of the profile initialization.

\begin{figure*}
\includegraphics[width=0.95\textwidth]{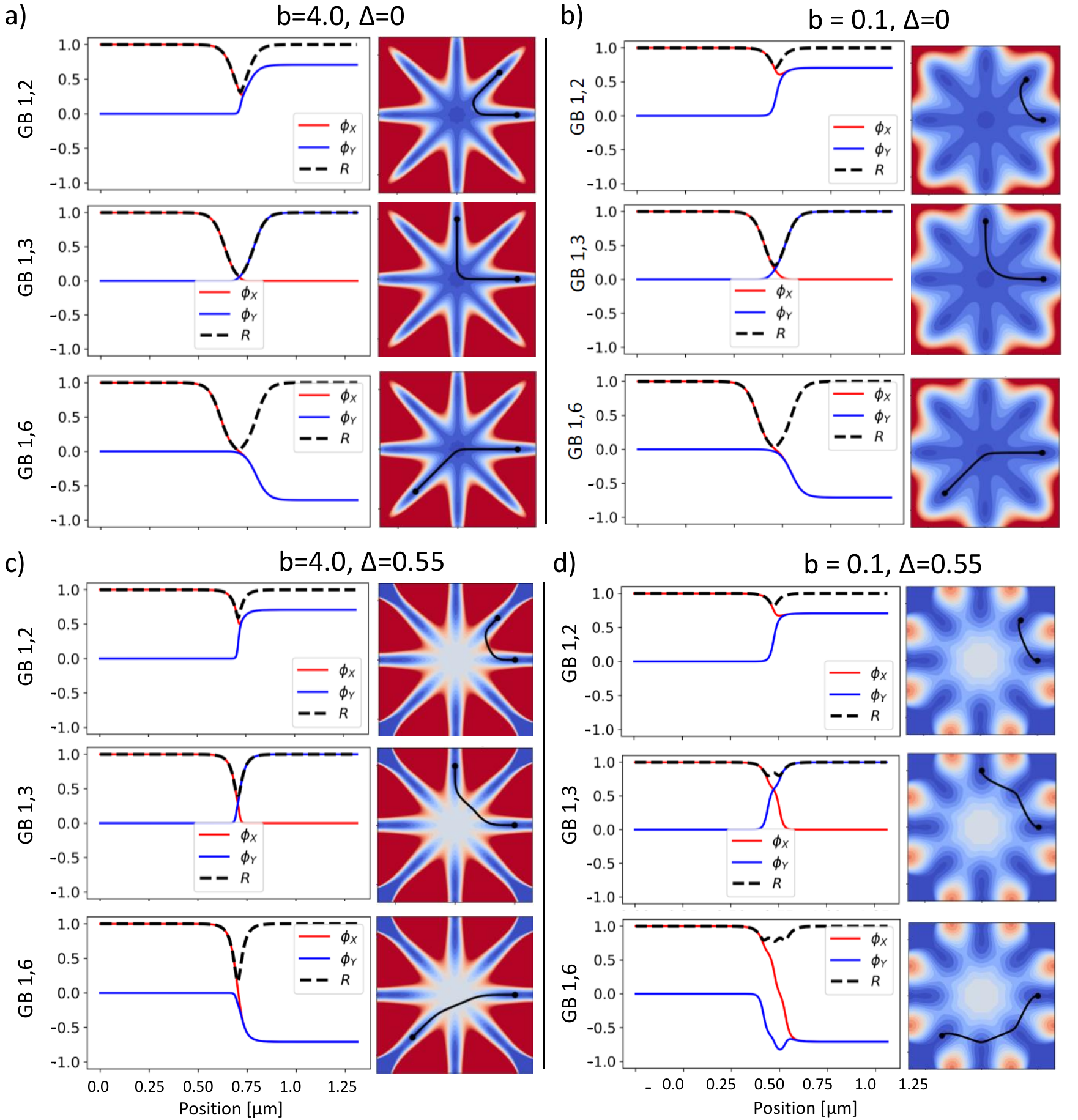}
\caption{Grain boundary profiles and vector order parameter trajectories in Landau landscapes with zero driving force $\Delta = 0$ (a-b)  and a relatively large driving force $\Delta = 0.55$ (c-d). Two different solid-solid barrier heights are considered ($b=0.1$ and $b=0.4$).  Concentration dynamics are not considered here.
}
\label{fig:GB_profiles}
\end{figure*}

The resulting profiles are shown in \cref{fig:GB_profiles} for a solid-solid barrier parameter $b = 4.0$ and $b=0.1$, along with the corresponding vector order parameter trajectories defining each grain boundary in the Landau free energy landscape (superimposed as a black solid line). For each choice of $b$, the driving force was set to either zero, i.e. $\Delta := (T_l - T)/((1-k_e) |m_l^e| c_0) = 0$, or $\Delta$ = 0.55. In all cases, solid-liquid ordering, represented by $R$, develops a typical cusp-like morphology in the grain boundary region consistent with earlier polycrystalline PF models, such as in orientation field models \cite{korbuly2017}. Moreover, the solid-solid grain boundary ordering decreases as the misorientation increases, approaching a ``wet'' grain boundary at high misorientation, as expected. The use of a smaller solid-solid barrier parameter $b=0.1$ increases the ordering at the grain boundary, which is consistent with a smaller energy barrier between the solid wells. Finally, applying a driving force ($\Delta$ = 0.55) increases the ordering at the grain boundary as expected.

In \cref{fig:GB_profiles}d, one can see slight oscillations in the solid-liquid ordering across the grain boundaries GB1,3 and GB1,6. This is expected for the small solid-solid barrier coefficient ($b$ = 0.1) and large quench ($\Delta$ = 0.55), as it is energetically favorable for the order parameter profile to approach the intermediate solid well, as shown in the grain boundary trajectories (black lines) in \cref{fig:GB_profiles}d. Curiously, this type of solid-liquid ordering oscillation is also found in phase field crystal simulations of Mellenthin et al \cite{Mellenthin08}. However, we expect that these solid-liquid ordering oscillations are uncommon for metals and should be generally avoided by setting the solid-solid barrier to a sufficiently large value, especially when the computational interface width $W_0$ is increased to thin interface conditions.

\begin{figure}
\includegraphics[width=0.49\textwidth]{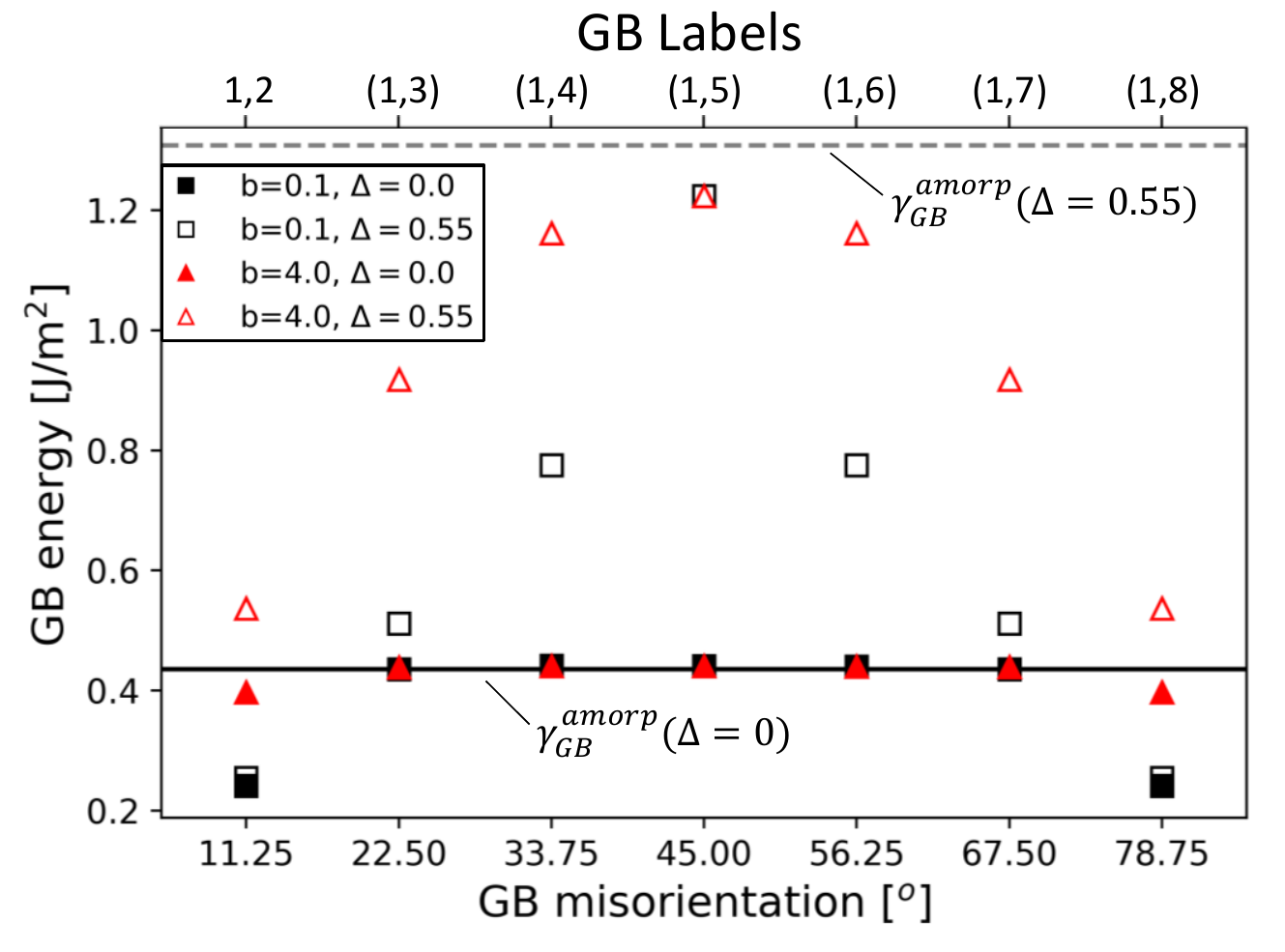}
\caption{Grain boundary energy versus grain boundary misorientations for the case of $N_{wells} = 8$.  Grain boundaries were grown with Model A dynamics (no concentration dynamics) for two values of undercooling,  $\Delta=$0 or 0.55. For each, two values of the solid-solid barrier parameter were used, $b=0.1$ or $b=4.0$. Analytical grain boundary energies, $\gamma_{GB}^{amorp}(\Delta)$, are shown as horizontal lines, where the solid line corresponds to $\Delta = 0$  and the dashed line to $\Delta = 0.55$.
}
\label{fig:GB_energies}
\end{figure}

We next evaluated grain boundary energies, the results of which are shown in \cref{fig:GB_energies}. The grain boundary energy is defined as the excess energy at the interface, which for 1D systems can be shown to have the form
\begin{align}
    \gamma_{GB} = \int_{\infty}^{\infty} &\Bigg( \frac{1}{2} \epsilon_0^2 \Big( \Big(\pxpy{}{x}\phi_X\Big)^2  + \Big(\pxpy{}{x}\phi_Y\Big)^2\Big)
    \Bigg. \nonumber \\ \Bigg. &
    + H\Big(  \, R^2 - \frac{1 + b\cos(N_{wells} \theta_\phi)}{1+b} R^3 + R^4  \Big)
    \Bigg. \nonumber \\ \Bigg. &
    + P(R) \Delta
    \Bigg) dx,
    \label{eq:GB_energy}
\end{align}
where
\begin{align}
   \frac{1}{H} &= a_1 \frac{W_0}{d_0} \frac{1}{c_0^2 g_{c,c}^\ell}, 
  \nonumber \\
    \epsilon_0 &= W_0 \sqrt{H},
    \nonumber \\
   \Delta &= \frac{T_l - T}{(1-k_e)|m_l^e| c_0}.
\end{align}
For an ideal amorphous grain boundary, an estimate of the energy can be derived analytically; the details are provided in the Supplemental material \cite{supplementalMaterial}. An ideal amorphous grain boundary assumes that the Landau angle $\theta$ is constant except at the center of the grain boundary, where the ordering goes to zero. Without a driving force ($\Delta = 0$), this energy has a closed-form expression that is twice the solid-liquid interface energy given by 
\begin{align}
\gamma_{GB}^{amorp} &= 2 \cdot \gamma_{SL} = \frac{W_0 H}{\sqrt{3}}
\approx 0.44 \text{ J/m}^2, 
\end{align}
where parameters are taken from \cref{table:phaseField_parameters}. When an undercooling (here $\Delta = 0.55$) is applied, the grain boundary energy  needs to be numerically computed, yielding $\gamma_{GB}^{amorp}(\Delta=0.55) \approx 1.31$ J/m$^2$ (see the Supplemental material \cite{supplementalMaterial} for more details). These estimated grain boundary energies ($\gamma_{GB}^{amorp}$) provide ideal limits for comparison and are shown as horizontal lines in \cref{fig:GB_energies}. These energies are compared for the Model A-equilibrated grain boundary profiles shown in \cref{fig:GB_profiles}, where the excess grain boundary energies are calculated by numerically evaluating \cref{eq:GB_energy}, and the resulting energies are shown by the scatter data in \cref{fig:GB_energies}.

The grain boundary energies obtained from our model are in good agreement with previous phase field crystal simulations of Mellenthin et al. \cite{Mellenthin08}, which are characterized by three main features. First, without a driving force (filled markers in \cref{fig:GB_energies}), the smallest misorientations (up to misorientation $\sim$ $11.25^o$), produce slightly attractive grain boundaries. Second, large misorientations up to the symmetry point for cubic crystals leads to  grain boundary energies identical to the ideal grain boundary energy, corresponding to zero applied driving force ($\gamma_{GB}^{amorp}(\Delta = 0$)). Third, adding a driving force of $\Delta = 0.55$ (hollow markers in \cref{fig:GB_energies}) increases the grain boundary energy in a gradual manner as a function of misorientation.

As expected, an increase in solid-solid barrier parameter $b$ leads to a corresponding increase in grain boundary energy for all cases, except for those cases where the energy has already saturated to it's limiting value, i.e. when the grain boundary has saturated to a large misorientation (for $\Delta$ = 0 misorientations 22.50-45$^o$ and its corresponding crystalline symmetric values, while for $\Delta$ = 0.55 similar saturation is not found). 

When we consider a thermodynamic driving force applied to the grain boundary, the simulated Model A energies at the maximum misorientation are smaller than our analytical estimate (dashed horizontal line in \cref{fig:GB_energies}). This discrepancy is due to the fact that in the analytical estimate, we assume that solid-liquid order is zero in the grain boundary region, and that $\theta$ varies only at zero ordering, neither of which are well satisfied for $\Delta$ = 0.55, particularly for small $b$ ($b$ = 0.1), which, as discussed above, leads to small interfacial oscillations within the interface. 

\subsubsection{Grain boundary back-diffusion}

This section demonstrates that the solute evolution, leading to the development of cohesive solid networks and grain boundaries, follows behavior consistent with the seminal of Rappaz and co-workers \cite{Rappaz03}. For this part of the study, we activate the concentration dynamics of \cref{eq:c_evolution}, and apply a constant cooling rate $2\times10^5$ K/s. In addition, we allow for enhanced back-diffusion by assuming that the solid diffusion coefficient is $D_s = D_\ell/100$. Here, the two solid regions were initialized with a thick liquid layer in between them to emulate, in 1D, the conditions of grain-grain merger during solidification.

\begin{figure*}
\includegraphics[width=0.9\textwidth]{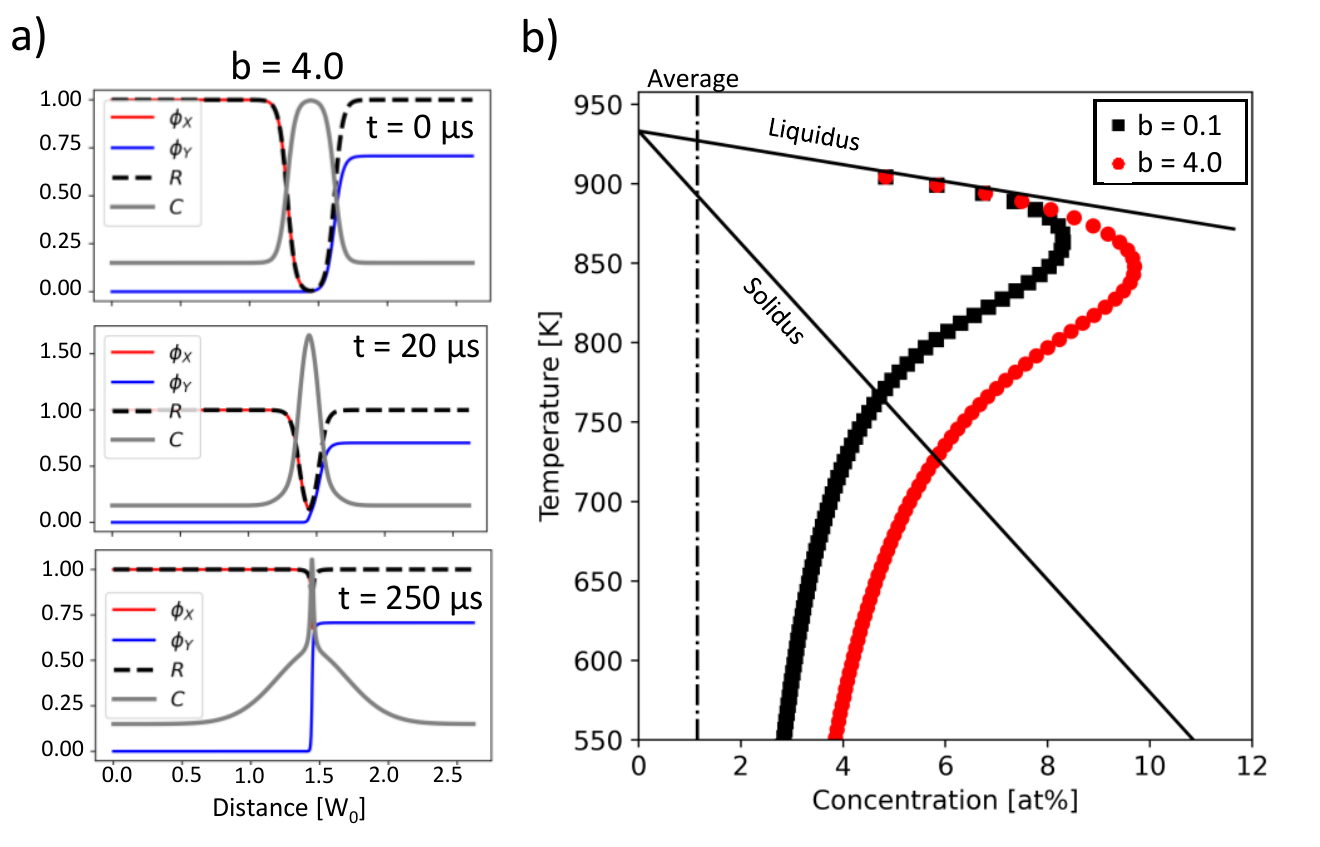}
\caption{Grain boundary coalescence for a cooling rate $10^6$ K/s, showing the evolution of the corresponding order parameter and concentration profiles. a) Solid-liquid ordering and concentration at three time steps for $b$=4.0, where concentration (C) is in units of the average composition $c_0$ = 4.5 at\%Cu; b) Grain boundary peak concentration at various time steps for $b=$0.1, and $b=$4.0. }
\label{fig:coalescence}
\end{figure*}

Back diffusion was simulated in this grain boundary system having the smallest misorientation allowed by our model (labeled GB 1,2 in \cref{fig:GB_energies}), corresponding to a misorientation of 11.25$^o$. The data was collected at two solid-solid barrier parameters, $b$ = 0.1 and $b$ = 4.0. The snapshots of the phase field profiles and the corresponding  concentration profiles are shown in \cref{fig:coalescence}a, where time increases from top to bottom. \cref{fig:coalescence}b plots the peak grain boundary concentration as a function of time superimposed on the dilute Al-Cu phase diagram. At early times, when the tails of the two grains are not in contact, the concentration follows the liquidus line. As the two grains begin to ``feel'' one another through their order parameter tails, and start to interact, the concentration starts to drop towards the solidus concentration. Due to back-diffusion, they asymptotically start to approach the average alloy concentration (vertical dashed line). 
This behavior is consistent with the back-diffusion results reported by Rappaz et al. \cite{Rappaz03} and Ofori-Opoku and Provatas \cite{Ofori-Opoku10}. Using a large solid-solid barrier parameter, $b$ = 4.0 (red circles in \cref{fig:coalescence}b), decreases the solid-solid ordering around the grain boundary for a given undercooling, and thereby increases the concentration peak, as expected. 

Overall, the above 1D grain boundary back-diffusion study  shows the expected behavior in terms of solid-solid ordering, grain boundary energy behavior, grain boundary merger and solute distribution at the grain boundary. It is also noteworthy that the properties reproduced by the proposed vector order parameter model have also all been exhibited by the more fundamental phase field crystal model, as documented in Ref.~\cite{Mellenthin08}, as well as in more traditional phase field models~\cite{Rappaz03,Ofori-Opoku10}.

\subsection{Polycrystalline grain coarsening} 
\label{section:results_modelA}

\begin{figure*}
\includegraphics[width=0.97\textwidth]{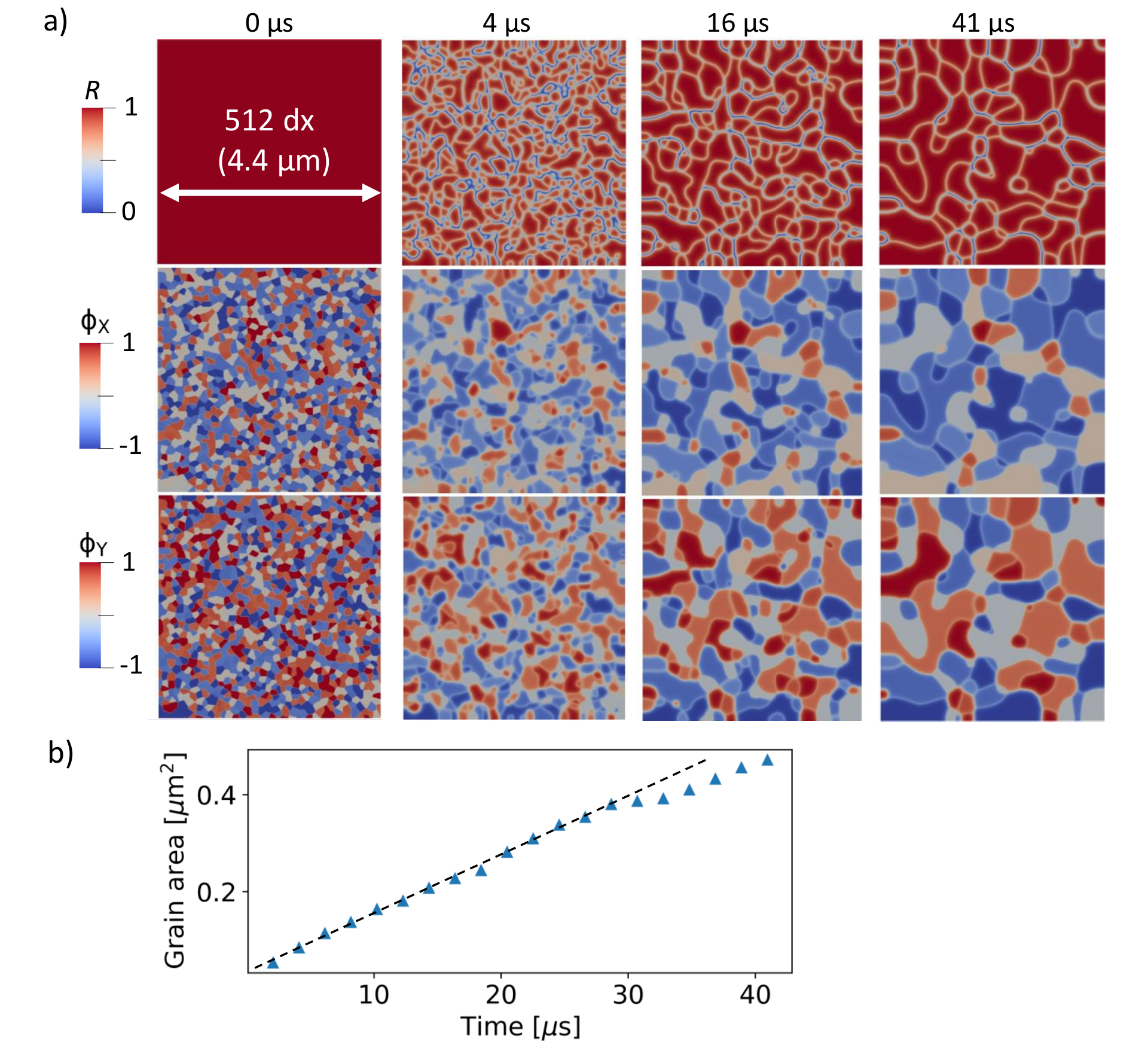}
\caption{Polycrystalline coarsening using the proposed vector order parameter model, constrained to Model A (no solute dynamics), with an undercooling of $\Delta = 0.55$ and solid-solid barrier $b=4.0$. a) Time evolution of polycrystalline grain network, where time increases from left to right. The rows show different measures of the order parameter; b) Solid fraction versus time, with the linear scaling regime is shown by the dashed line as a guide for the eye.  
}
\label{fig:model_A_coarsening}
\end{figure*}

In this section, we verify that the Model A dynamics of the proposed vector order parameter model satisfies the theoretically established $\sim t^{1/2}$ coarsening kinetics.  For this study, we set the grid size $\Delta x = 0.2 W_0$ as we found that larger value of mesh size (e.g., $\Delta x = 0.5W_0$) can lead to lattice pinning, especially when a large driving force is applied. Simulations were performed using a driving force  $\Delta = 0.55$, an interaction barrier strength of $b=4.0$, and we applied zero gradient boundary conditions. For this study, we initialized the simulation domain with a random Voronoi tessellation with full ordering $R=\phi_X^2 + \phi_Y^2 = 1$ within the bulk of each grain, as shown in the first column of \cref{fig:model_A_coarsening}a. The other columns of \cref{fig:model_A_coarsening}a show the time evolution and grain coarsening in the polycrystalline system, where each row highlights different measures of the vector order parameter. It is noteworthy that the regions corresponding to high-angle grain boundaries lead to the emergence of solid-liquid ordering. 

To verify that the grain coarsening follows the well-known results of linear (parabolic) growth with area (radius) versus time, we evaluated the average grain size using the so-called intercept method, which is typically used in experimental settings to determine grain sizes. In this method, a line is drawn randomly across the microstructure and the size of the grains intercepting this line is measured. A sufficiently large number of these line-based measurements, at various orientations through the microstructure data, leads to a good estimate of the grain size $d_{grain}$. We converted the so-calculated average grain size to an equivalent circle via $\pi (d_{grain}/2)^2$. The grain area versus time is shown in \cref{fig:model_A_coarsening}b. Mean-field arguments, stating that grains grow by mean curvature, dictate that the grains should coarsen such that the average grain area increases linearly with time. This linear coarsening regime can be seen in \cref{fig:model_A_coarsening}b, where a dashed line is drawn as a guide for the eye for the linear regime. As the grain size approaches the system size, the mean field assumption breaks down and the scaling relation no longer holds.

\subsection{Nucleation and polycrystalline solidification in an alloy} 
\label{section:results_modelC}

This section demonstrates noise-induced  nucleation and polycrystalline solidification in a binary alloy using the proposed vector order parameter model.  The aim here is to highlight the robustness of the model and the self-consistency of applying thermal noise acting on the vector order parameter defined everywhere. 
To prevent interference with the evolution of actual ordered crystals, we threshold the noise terms such that it is only active in the vicinity where $R = \sqrt{\phi_X^2 + \phi_Y^2} < 0.9$. We additionally filter the noise in space and time, so that the fluctuations occur roughly over a length scale of $W_0$ (not $\Delta x$), and over the time scale $\tau_0$ (not $\Delta t$). The details of these filtering procedures are given in the Supplemental material \cite{supplementalMaterial}. These filtering procedures were necessary to ensure the numerical stability of the system, and fluctuations below the length and time scales ($W_0$ and $\tau_0$) have no clear physical meaning in coarse grained field theories. For the simulations shown in this section we used periodic boundary conditions and a cooling rate of $1.5\times 10^5$ K/s, and initialized the system as a uniform liquid.  

\begin{figure*}
\includegraphics[width=0.97\textwidth]{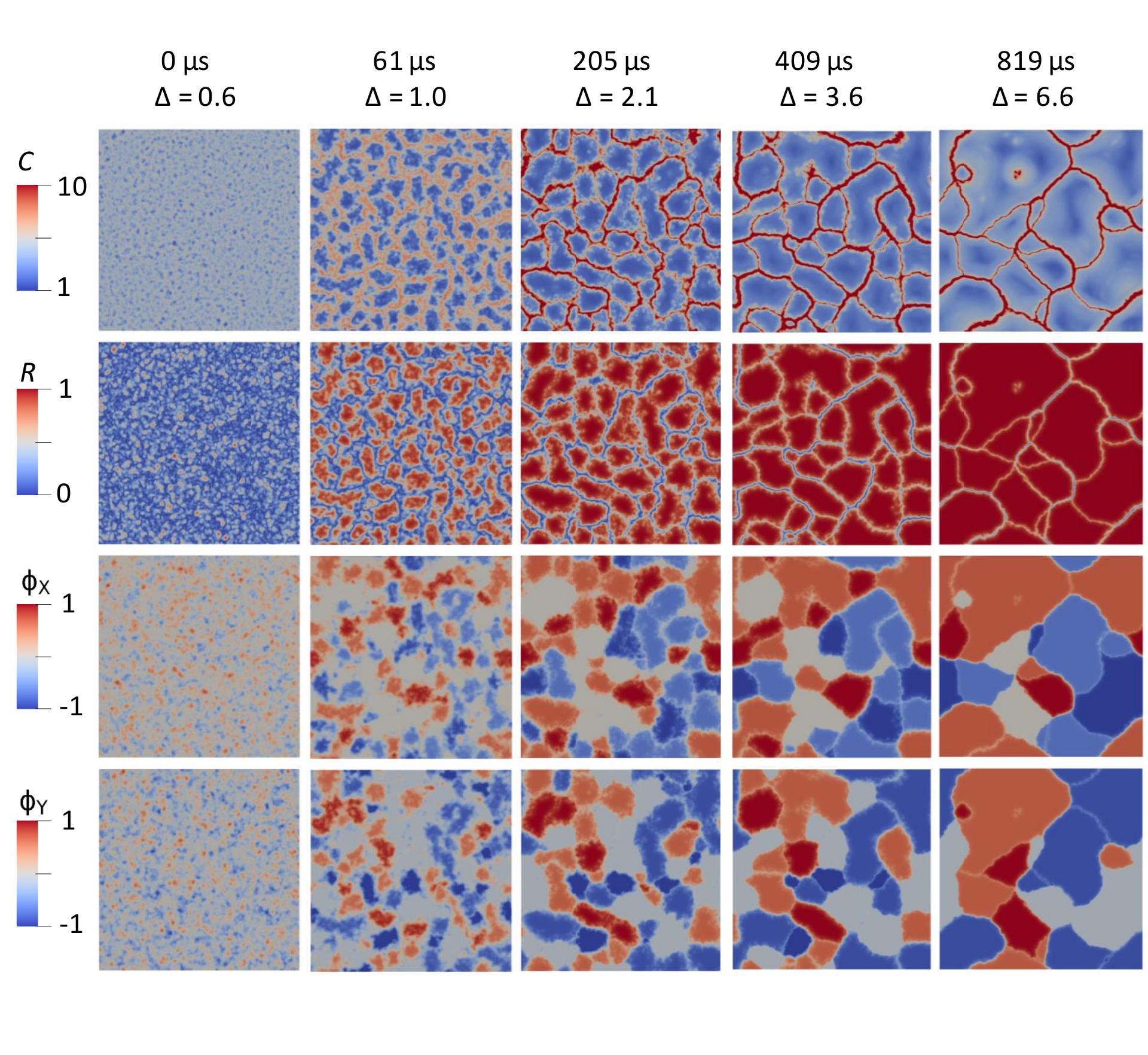}
\caption{Thermally induced homogeneous nucleation, polycrystalline solidification and coarsening in a binary alloy simulated  with the proposed vector order parameter phase field model.  Model parameters are taken from \Cref{table:phaseField_parameters} ($\Delta x = 0.2 W_0$). The cooling rate is set to $1.5\times10^5$ K/s. Time flows from left to right as the corresponding undercooling increases. The rows indicate the fields of the model.}
\label{fig:model_C_nucleation}
\end{figure*}

\cref{fig:model_C_nucleation} shows the nucleation and subsequent time evolution of a polycrystalline network of grains in a binary alloy, starting from a liquid state and with thermal fluctuations applied over time. Time goes from left to right as the corresponding undercooling ($\Delta$) grows. The rows display the concentration field and different measures of the vector order parameter. The data shows that  
the thermal fluctuations grow over time as the driving force increases, leading to homogeneous nucleation of globular structures, i.e. the nuclei, as shown in the first column of  \cref{fig:model_C_nucleation}. The nuclei then start to merge and coalesce into a polycrystalline network, which starts to coarsen over time. The grain boundaries in this set of simulations appear jagged and noisy due to thermal fluctuations, but eventually do smooth out (coarsen) due to surface energy minimization at later times. 

\section{Discussion}
\label{sec:discussions}

In the proposed vector order parameter phase field model, the number of solid wells, i.e. the number of unique grains, is practically limited to a value of roughly $N_{wells}$ = 10, however this can be increased easily if the grid spacing $\Delta x$ is decrease. A more practical and simple technique for increasing the number of grains that avoids the excessive computational cost associated with decreasing $\Delta x$ would be to increase the vector order parameter dimensionality from $d=$2 ($\phi_X$ and $\phi_Y$) to e.g. $d$ = 3 or 4 ($\phi_K$, $K$ = 1, ..., $d$). 

The  solid wells in the energy landscape defined by the vector order parameter can also be used to represent multiple thermodynamically distinct solid phases. Here the aforementioned increase of vector order parameter dimensionality could be used to assign different solid wells to different phases. In our preliminary tests using the local supersaturation approximation \cite{Plapp11,raj2017}, the number of solid wells ($N_{wells}$) needs to be rather small to describe phases with notably different solubilities, such as an $\alpha$ FCC aluminium phase and an intermetallic Al$_2$Cu phase. 

In order to reduce computational costs associated with the use of multiple order parameters to represent distinct phases, multi-phase field model or multi-order parameter phase field models can use so-called re-indexing algorithms as a way to re-assign grains to avoid artificial melding of neighboring grains \cite{krill02}. This re-indexing might not be possible for the current vector order parameter model, as the order parameters of neighboring grains are directly interacting with each other and relax to profiles such as those shown in \cref{fig:GB_profiles}. However, the use of higher dimensional vector order parameter fields, coupled with adaptive mesh refinement should be able to provide ample degrees of freedom to simulate polycrystalline networks relatively efficiently in the context of a single physical order parameter.

The anisotropy of grain boundaries can be an important feature in polycrystalline coarsening, characterized in 3D by three misorientation angles and two inclination angles \cite{rohrer2011}---in a 2D system by two misorientation angles and a single inclination angle. In its current form, the present model depends mostly on the misorientation, i.e. on which solid wells are in contact. There is a minor inclination dependence through the anisotropic solid-liquid interface energy.  We expect that the inclination dependence can be increased by, for example, adding an anisotropic energy term proportional to $|\theta|^2$ for the energy expression in \cref{eq:totalFreeEnergyFunctional},  analogously to how it is done for the solid-liquid interface energy.

Physically, grain boundaries move significantly slower than solid-liquid interface. As grain boundaries are necessarily associated with a change in orientation, the mobility decrease could be implemented by adding an orientation gradient $|\nabla\theta|^2$ dependence to the phase field time scale, for example in the form $\tau_K \rightarrow (1-\exp(-A |\nabla \theta|^2))\tau_K$, where $A$ controls how strongly the mobility decays around grain boundaries. 

Numerically, the present vector order parameter model requires a relatively dense grid, with roughly $\Delta x \sim 0.2 W_0$, in order to resolve the grain boundary profiles and to avoid mesh pinning. This becomes more arduous when the driving force or undercooling is increased since this leads to steeper profiles, a feature also present to some extent in all phase field models. This requires further investigation to find possible numerical remedies.

The vector order parameter model can be more elegantly formulated in polar coordinates $R, \theta$ (see \cref{fig:landau_landscape}). However, in our tests, the system is prone to mesh pinning due to topological defects analogical to those seen by Korbuly et al. \cite{korbuly2017} for the orientation phase field model. For those readers interested in purusing this formulation, Korbuly et al. present techniques and remedies to address these issues that are expected to also be applicable to the present model.

\section{Conclusion}
\label{sec:conc}
The vector order parameter phase field model presented in this work is introduced as a practical tool for the study of microstructure evolution in polycrystalline solidification, and which reproduces key benchmarks quantitatively. The specific benchmarks that were demonstrated in this work are:  dendrite solidification \cite{Karma01,Echebarria04}, grain boundary energy versus misorientation and undercooling, coalescence and segregation associated with back-diffusion and nucleation-induced solidification and coarsening in binary alloys. 

Future work with this new phase field model will explore the addition of higher dimensional vector order parameter field in order to increase the efficiency increasing the number of unique grains, and to introduce new thermodynamically distinct phases, thus making possible the efficient simulation of multi-phase solidification, such as for example the simulation of  intermetallic phases in the Al-Cu system.

\section*{Acknowledgement}
NP acknowledges the National Science and Engineering Research Council of Canada (NSERC) and the Canada Research Chairs (CRD) Program. TP and AL acknowledge the support of Academy of Finland through the HEADFORE project, Grant No. 333226.
\appendix

\section{ Variational derivative of interface energy term}
This section shows the explicit expression for the interface energy term for the vector order parameter model, \cref{eq:fint}.
Let's first separate the anisotropy function from interface width as $W_K = W_0 \,a_K$.
Then the gradient penalty free energy contribution for order parameter model is
\begin{align}
\int_V f_{int} dV = \int_V  \sum_{K =X,Y} W_0^2 \frac{1}{2}|a_{K}(\theta_{\nabla \phi_K}, \theta) \nabla \phi_K |^2 dV,
\end{align}
where 
\begin{align}
    a_{K}( \phi_X, \phi_Y, \nabla \phi_K)  =  1 + \epsilon_c \cos \left(4 \theta_{\nabla \phi_K} - \theta \right) ,
\end{align}
where the local interface normal angle is 
\begin{align}
    \theta_{\nabla \phi_K} = \arctan \left( \frac{ \partial_y \phi_K}{\partial_x \phi_K} \right),
\end{align}
and the reference angle of order parameter vector $\vec{\phi} = (\phi_X, \phi_Y)$ is
\begin{align}
    \theta = \arctan \left( \frac{ \phi_Y}{\phi_X} \right).
\end{align}
For order parameter vector components $K \in \lbrace X,Y \rbrace$, the total variational derivative of this term is then
\begin{align}
&\frac{\delta}{\delta \phi_{K} }\left( \sum_{K'=X,Y} \frac{1}{2}|a_{K'} \nabla \phi_{K'}|^2 \right) \nonumber \\
=&
\left( 
\pxpy{}{\phi_K} - \nabla \cdot \pxpy{}{\nabla \phi_K} 
\right)
\left( \sum_{K'=X,Y} \frac{1}{2}|a_{K'} \nabla \phi_{K'}|^2 \right)
\nonumber \\
= &
a_{X} \pxpy{a_{X}}{\phi_K} |\nabla \phi_{X}|^2
+
a_{Y} \pxpy{a_{Y}}{\phi_K} |\nabla \phi_{Y}|^2
\nonumber \\ 
& - \nabla \cdot \left(
a_{K}^2 \nabla \phi_K  
+ a_{K} |\nabla \phi_K|^2 \pxpy{a_{K}}{\nabla \phi_K}
\right),
\end{align}

where $\pxpy{}{\nabla \phi_K} := \sum_{i}^{x,y,z} \hat{e}_i \pxpy{}{_i (\partial \phi_K)}$, and the following identities are used:
\begin{align}
 \pxpy{\theta}{\phi_X} &= -\frac{\phi_Y}{|\phi|^2}
 \text{, }\;\;\;\;
 \pxpy{\theta}{\phi_Y} = \frac{\phi_X}{|\phi|^2},
 \\
    \pxpy{\theta_{\nabla \phi_K}}{(\partial_x \phi_K)} &= -\frac{ \partial_y \phi_K }{|\nabla \phi_K|^,2}
    \text{, }\;\;\;\;
    \pxpy{\theta_{\nabla \phi_K}}{(\partial_x \phi_K)} = \frac{ \partial_x \phi_K }{|\nabla \phi_K|^2},
    \\
    a_{K}'(\theta_{\nabla \phi_K}, \theta) &= -4 \epsilon_c \sin( 4\theta_{\nabla \phi_K}-\theta),
    \\
    \pxpy{a_{K}}{ (\partial_x \phi_K)} &= a_{K}'(\theta_{\nabla \phi_K}, \theta) \pxpy{\theta_{\nabla \phi_K}}{(\partial_x \phi_K)} 
    \nonumber\\ 
    &= -a_{K}'(\theta_{\nabla \phi_K}, \theta) \frac{ \partial_y \phi_K }{|\nabla \phi_K|^2},
    \\
    \pxpy{a_{K}}{(\partial_y \phi_K)} &= a_{K}'(\theta_{\nabla \phi_K}, \theta) \pxpy{\theta_{\nabla \phi_K}}{(\partial_y \phi_K)}  
    \nonumber \\ 
    &= a_{K}'(\theta_{\nabla \phi_K}, \theta) \frac{ \partial_x \phi_K }{|\nabla \phi_K|^2},
\\
    \pxpy{a_{K}}{\phi_K} & 
     =\epsilon_c \sin\left( 4 \left( \theta_{\nabla \phi_K} - \theta/4 \right) \right) \pxpy{\theta}{\phi_K}
    \nonumber \\ 
    &= -\frac{1}{4} a_{K}'(\theta_{\nabla \phi_K}, \theta) \pxpy{\theta}{\phi_K}.
\end{align}
%

%

\begin{thebibliography}{10}
\expandafter\ifx\csname url\endcsname\relax
  \def\url#1{\texttt{#1}}\fi
\expandafter\ifx\csname urlprefix\endcsname\relax\def\urlprefix{URL }\fi
\expandafter\ifx\csname href\endcsname\relax
  \def\href#1#2{#2} \def\path#1{#1}\fi

\bibitem{Ste99}
I.~Steinbach, F.~Pezzolla, Physica D 134 (1996) 385.

\bibitem{Bottger06}
B.~Bottger, J.~Eiken, I.~Steinbach, Acta Materialia 54 (2006) 2697.

\bibitem{Ofori-Opoku10}
N.~Ofori-Opoku, N.~Provatas, A quantitative multi-phase field model of
  polycrystalline alloy solidification, Acta Materialia 58~(6) (2010) 2155 --
  2164.

\bibitem{kobayashi1998}
R.~Kobayashi, J.~A. Warren, W.~C. Carter, Vector-valued phase field model for
  crystallization and grain boundary formation, Physica D: Nonlinear Phenomena
  119~(3-4) (1998) 415--423.

\bibitem{granasy2002}
L.~Gr{\'a}n{\'a}sy, T.~B{\"o}rzs{\"o}nyi, T.~Pusztai, Crystal nucleation and
  growth in binary phase-field theory, Journal of Crystal Growth 237 (2002)
  1813--1817.

\bibitem{Warren03}
J.~A. Warren, R.~Kobayashi, A.~E. Lobkovsky, W.~C. Carter, Acta Materialia 51
  (2003) 6035.

\bibitem{korbuly2017}
B.~Korbuly, M.~Plapp, H.~Henry, J.~A. Warren, L.~Gr{\'a}n{\'a}sy, T.~Pusztai,
  Topological defects in two-dimensional orientation-field models for grain
  growth, Physical Review E 96~(5) (2017) 052802.

\bibitem{pusztai2005}
T.~Pusztai, G.~Bortel, L.~Gr{\'a}n{\'a}sy, Phase field theory of
  polycrystalline solidification in three dimensions, EPL (Europhysics Letters)
  71~(1) (2005) 131.

\bibitem{steinbach1999}
I.~Steinbach, F.~Pezzolla, A generalized field method for multiphase
  transformations using interface fields, Physica D: Nonlinear Phenomena
  134~(4) (1999) 385--393.

\bibitem{salama2020}
H.~Salama, J.~Kundin, O.~Shchyglo, V.~Mohles, K.~Marquardt, I.~Steinbach, Role
  of inclination dependence of grain boundary energy on the microstructure
  evolution during grain growth, Acta Materialia (2020).

\bibitem{krill02}
C.~E.~K. III, L.-Q. Chen, Acta Materialia 50 (2002) 3057.

\bibitem{morin1995}
B.~Morin, K.~R. Elder, M.~Sutton, M.~Grant, Model of the kinetics of
  polymorphous crystallization, Physical Review Letters 75~(11) (1995) 2156.

\bibitem{Plapp11}
M.~Plapp, Unified derivation of phase-field models for alloy solidification
  from a grand-potential functional, Physical Review E 84~(3) (2011) 031601.

\bibitem{ProvatasElder10}
N.~Provatas, K.~Elder, Phase-Field Methods in Materials Science and
  Engineering, Wiley-VCH Verlag GmbH \& Co. KGaA, 2010.
\newblock \href {https://doi.org/10.1002/9783527631520.ch1}
  {\path{doi:10.1002/9783527631520.ch1}}.

\bibitem{WBM1992}
W.~J.~B. A.~A.~Wheeler, G.~B. McFadden, Physical Review A 45 (1992) 7424--7439.

\bibitem{Karma01}
A.~Karma, Phase-field formulation for quantitative modeling of alloy
  solidification, Phys. Rev. Lett 87 (2001) 115701.

\bibitem{Echebarria04}
B.~Echebarria, R.~Folch, A.~Karma, M.~Plapp, Quantitative phase-field model of
  alloy solidification, Physical Review E 70 (2004) 061604--1.

\bibitem{Karma98}
A.~Karma, W.~J. Rappel, Quantitative phase-field modeling of dendritic growth
  in two and three dimensions, Physical Review E 57 (1998) 4323--4349.

\bibitem{Almgren99}
R.~Almgren, SIAM J. Appl. Math. 59 (1999) 2086.

\bibitem{karma1999}
A.~Karma, W.-J. Rappel, Phase-field model of dendritic sidebranching with
  thermal noise, Physical Review E 60~(4) (1999) 3614.

\bibitem{supplementalMaterial}
See supplemental material at [url will be inserted by publisher] for a finite
  volume expression of the interface energy penalty, analytical grain boundary
  energy estimates, spatial and temporal filtering of the thermal noise, and
  polycrystalline coarsening using a small solid-solid barrier parameter.

\bibitem{Greenwood18}
M.~Greenwood, K.~Shampur, N.~Ofori-Opoku, T.~Pinomaa, L.~Wang, S.~Gurevich,
  N.~Provatas, Quantitative 3d phase field modelling of solidification using
  next-generation adaptive mesh refinement, Computational Materials Science 142
  (2018) 153.

\bibitem{Mellenthin08}
J.~Mellenthin, A.~Karma, M.~Plapp, Phase-field crystal study of grain-boundary
  premelting, Physical Review B 78~(18) (2008) 184110.

\bibitem{Rappaz03}
M.~Rappaz, A.~Jacot, W.~J. Boettinger, Metallurgical and Materials Transactions
  A 34~(3) (2003) 467--479.

\bibitem{raj2017}
K.~Shampur, A grand potential based multi-phase field model for alloy
  solidification, Master's thesis, McGill University (2017).

\bibitem{rohrer2011}
G.~S. Rohrer, Grain boundary energy anisotropy: a review, Journal of Materials
  Science 46~(18) (2011) 5881--5895.

\end{thebibliography}

\end{document}